# Reaction-Diffusion Waves Coupled with Membrane Curvature

*Naoki Tamemoto and Hiroshi Noguchi*[*]

Institute for Solid State Physics, University of Tokyo, Kashiwa, Chiba 277-8581, Japan

**Abstract**

The reaction-diffusion waves of proteins are known to be involved in fundamental cellular functions, such as cell migration, cell division, and vesicular transportation. In some of these phenomena, pattern formation on the membranes is induced by the coupling between membrane deformation and the reaction-diffusion system through curvature-inducing proteins that bend the biological membranes. Although the membrane shape and the dynamics of the curvature-inducing proteins affect each other in these systems, the effect of such mechanochemical feedback loops on the waves has not been studied in detail. In this study, reaction-diffusion waves coupled with membrane deformation are investigated using simulations combining a dynamically triangulated membrane model with the Brusselator model extended to include the effect of membrane curvature. It is found that the propagating wave patterns change into nonpropageting patterns and spiral wave patterns due to the mechanochemical effects. Moreover, the wave speed is positively or negatively correlated with the local membrane curvature depending on the spontaneous curvature and bending rigidity. In addition, self-oscillation of the vesicle shape occurs, associated with the reaction-diffusion waves of curvature-inducing proteins. This agrees with the experimental observation of liposomes with a reconstituted Min system, which plays a key role in the cell division of Escherichia coli. The findings of this study demonstrate the importance of mechanochemical coupling in biological phenomena.

## Introduction

In living cells, the reaction-diffusion waves of proteins are involved in basic biological phenomena such as cell migration[1-7], cell division,[8-11] and vesicular transportation[12-14]. As these waves on biological membranes are associated with the spatiotemporal dynamics of the cell cortex, actin dynamics have been extensively studied to understand the mechanism of such phenomena[15, 16]. For example, waves of actin and the related proteins are observed in the migration of various types of cells such as amoeboid cells[1-5], neutrophils[6], epithelial cells[7], and others[15]. These waves are self-organized in the intracellular reaction networks and their activities regulate cell morphology (e.g., protrusion of pseudopodia and lamellipodia); hence, they control the migratory mode and direction of cells. In cell division, the reaction-diffusion waves of Min proteins in *Escherichia coli* determine the division site by forming a ring-like structure of FtsZ protein[8, 17]. It has been reported that the waves in eukaryotic cells also associated with cell division have size-scaling properties, predict the division site,[10] and mediate cell division[11]. In addition, the protein waves on membranes are associated with vesicular transportation such as endocytosis[12] and macropinocytosis[14]. It has been demonstrated that collective waves of endocytic proteins are generated on basal cellular membranes, inducing endocytosis at hotspots. In macropinocytosis, the characteristic reaction-diffusion wave dynamics of actin on dorsal cellular membranes have been experimentally observed and analyzed through simulation[18, 19]. As mentioned above, various studies have shown that cortical waves play a fundamental role in cellular functions.

Recently, it has been reported that certain waves are accompanied by the concentration oscillation of curvature-inducing/sensing proteins, and therefore, the mechanochemical feedback loops between the reaction-diffusion systems and membrane deformation affect the wave properties[13, 19, 20]. In these systems, feedback from the membrane curvature on the reaction-diffusion waves can affect their patterns and properties. The participation of curvature-inducing proteins in cyclic cortical events has also been observed in migratory cells, and it has been suggested that they control the cell polarity[21], migratory mode, and direction[22]. Moreover, it has been reported that a reconstituted Min system in giant unilamellar vesicles (GUVs) spontaneously generates reaction-diffusion waves accompanying cyclic vesicle deformation[23-25]. However, although reaction-diffusion waves coupled with membrane curvature have been studied, the effect of the mechanochemical feedback loop on the waves is yet to be determined.

In view of the above, this study investigates reaction-diffusion waves coupled with membrane deformation through simulation. Although the coupling of membrane deformation and actin polymerization has been extensively studied, actin filaments can flatten the membrane but also bend it, depending on their interaction with the membrane. In contrast, the role of curvature-inducing proteins, such as F-BAR proteins[21, 26, 27], is well expressed as local curvature generation. Here, we consider the reaction of two proteins: one protein induces an isotropic spontaneous curvature and its concentration is regulated by the other. We analyze the coupling effects between the change in membrane curvature and the reaction-diffusion dynamics of the curvature-inducing protein. In our previous study[28], we had primarily focused on Turing patterns; hence, in this study, we focus on propagating waves.

## Model and methods

### Membrane model

In this study, we apply the model used in our previous study,[28] with a minor modification in the reaction. A membrane vesicle contains $N$ vertices connected by bonds of average length $a$, with excluded volumes and masses $m$. The connected bond network creates a triangulated spherical surface, whose surface area $S$ and vesicle volume $V$ are constrained by harmonic potentials[29]. During the simulation, a diagonal bond is removed within two adjacent triangles including four vertices and five bonds, and the remaining two vertices are newly connected through the Monte Carlo method for modeling the membrane fluidity (bond-flip)[30].

In this model, the curvature free energy is represented as $F_{\mathrm{cv}} = \int f_{\mathrm{cv}} dS$, where $f_{\mathrm{cv}}$ is the local curvature energy:

$$f_{\mathrm{cv}} = (1-u)\frac{\kappa_0}{2}(2H)^2 + u\frac{\kappa_1}{2}(2H - C_0)^2, \qquad (1)$$

where $u$ is the concentration of the curvature-inducing proteins on the membranes ($u \in [0,1]$); $\kappa_1$ and $\kappa_0$ are bending rigidity with and without the proteins on the membranes, respectively; $C_0$ is the spontaneous curvature induced by the binding of the curvature-inducing proteins; $H$ is the mean curvature ($H = (C_1 + C_2)/2$, where $C_1$ and $C_2$ are two principal curvatures). Therefore, vesicles are deformed depending on the local spontaneous curvature, which changes based on the protein concentration $u$ and the spontaneous curvature $C_0$. Note that direct interactions between proteins are not considered here.

### Reaction-diffusion model

As a two-dimensional reaction-diffusion model, we employ the Brusselator model with a modified unbinding process for the curvature-inducing proteins (Fig.1a). The rate of the unbinding process is set to $k_{\text{ub}} = u/(1-u)$ such that $u \leq 1$. Thus, the reaction-diffusion equations are given by $\tau \partial u/\partial t = f(u,v) + D_u \nabla^2 u$ and $\tau \partial v/\partial t = g(u,v) + D_v \nabla^2 v$ with

$$f(u,v) = A - \left(B + \frac{u}{1-u}\right)u + u^2 v, \quad (2)$$

$$g(u,v) = Bu - u^2 v, \quad (3)$$

where $\tau$ is the time constant, $D_x$ is the diffusion coefficient of chemical reactant $x$, $\nabla^2$ is the two-dimensional Laplace–Beltrami operator, and $A$ and $B$ are positive parameters. By modifying the second term in $f(u,v)$, $u$ is maintained at $[0,1]$ (Fig. S1a). The linear stability analysis around the fixed point is presented in the ESI (Fig. S1b).

**Coupling of the membrane deformation and the reaction-diffusion model**

To couple the reaction-diffusion model with the membrane curvature, we assume that the membrane curvature affects only the binding processes of the protein $u$ through the local curvature energy $f_{\text{cv}}$ given in eqn (1). Because the free energy barrier can be changed by the membrane curvature, the reaction rate is not uniquely determined by the condition of the detailed balance[31]. Here, we consider that the barrier changes by the same amount as the bound state, as shown in Fig. 1b. Other choices, such as the linear dependence of the binding rate on the free energy change in ref. 28 and the Glauber rates in ref. 31, are also available. The entire reaction-diffusion equation is expressed as

$$\tau \frac{\partial u}{\partial t} = K_{\text{cv}} A - \left(B + \frac{u}{1-u}\right)u + u^2 v + D_u \nabla^2 u, \quad (4)$$

$$\tau \frac{\partial v}{\partial t} = Bu - u^2 v + D_v \nabla^2 v, \quad (5)$$

where $K_{\text{cv}} = \exp(-(1/k_{\text{B}}T) \partial f_{\text{cv}}/\partial u)$. Note that $\partial f_{\text{cv}}/\partial u$ is the internal energy difference for binding/unbinding, and the mixing entropy of proteins is not included, although this energy implicitly includes the entropy of the hidden degrees of freedom such as the conformational entropy of lipid molecules. The fixed point of the reaction equations $(u_s, v_s)$ is

$$u_s = \frac{2}{1 + \sqrt{1 + \frac{4}{K_{\text{cv}} A}}}, \quad v_s = \frac{B}{u_s}.$$

**Parameters**

In this study, we use membrane vesicles in which the number of vertices $N = 1006, 2004, 4000$, and $15994$. For these numbers of vertices, the vesicle surface areas are $S \simeq 823a^2, 1642a^2, 3278a^2$, and $13113a^2$, and the corresponding radii are $R = \sqrt{S/4\pi} \simeq 8.09a, 11.4a, 16.2a$, and $32.3a$, respectively. Unless otherwise mentioned, we applied parameters $A = 1, B = 13.2, D_u = 4, \eta = \sqrt{D_u/D_v} = 1, \tau = 10$ and $\kappa_0/k_{\text{B}} T = 20$ in all the simulations. The initial protein concentrations are set to $(u_s + w, v_s + w)$, where $u_s$ and $v_s$ are the fixed point of eqns (4) and (5) with a fixed value of the membrane curvature $H = 1/R$, and $w$ is a random number obtained from a Gaussian distribution function with a standard deviation of $0.01$. The other parameters values are the same as those used in ref. 28.

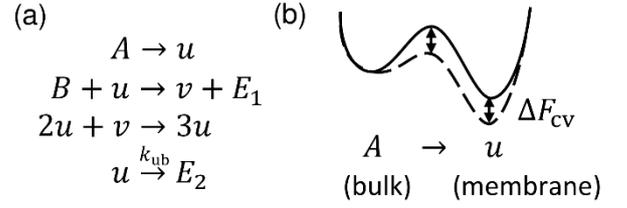

**Figure 1** Schematic model of the reaction-diffusion system employed in this study: (a) Brusselator model involving a modified unbinding process. In a normal Brusselator, $k_{\text{ub}} = 1$. (b) Concept for coupling of membrane curvature and reaction-diffusion system.

Equations (4) and (5) are numerically integrated using the finite-volume scheme[28]. Membrane deformation is solved through molecular dynamics simulation using a Langevin thermostat:

$$m \frac{\partial^2 \boldsymbol{r}_i}{\partial t^2} = -\frac{\partial U}{\partial \boldsymbol{r}_i} - \zeta \frac{\partial \boldsymbol{r}_i}{\partial t} + \boldsymbol{g}_i(t),$$

where $\zeta$ is the friction coefficient and $\boldsymbol{g}_i(t)$ is Gaussian white noise, which obeys the fluctuation-dissipation theorem. The potential $U$ is described in the ESI, which is the same one used in ref. 28. The simulation method is detailed in ref. 28. The error bars are calculated based on three independent runs.

To remove the effect of thermal fluctuations in the estimation of a local curvature, smoothed local curvature $\widetilde{H}$ is calculated by averaging the local curvature $H$ up to the second-order adjacent vertices (detailed in the Supplementary Material of ref. 28).

**Results and discussion**

**Spontaneous wave pattern generation**

We first investigated the pattern formation on deformable vesicles for $N = 15994$ at reduced volumes $V^* = 3V/4\pi R^3 = 0.65, 0.8$, and $0.95$. In the absence of protein binding, for $V^* \gtrsim 0.65$, the vesicles form prolate shapes at thermal equilibrium; for $0.59 \lesssim V^* \lesssim 0.65$, the shape at thermal equilibrium is a discocyte (biconcave disk) and the prolate is metastable[29, 32]. For $C_0 = 0$ and $\kappa_1/\kappa_0 = 1$, there is no curvature effect on the reaction-diffusion system because $K_{\text{cv}} = 1$ in eqn (4), and spatially homogeneous oscillations alone are observed in the protein concentration (purple squares in Fig. 2a).

Even when a propagating wave is initially induced by spatially inhomogeneous noise in the protein concentrations, it relaxes to a temporal oscillation uniformly on the entire vesicle surface.

In the presence of mechanochemical coupling, according to the nonuniform curvature of the vesicle, propagating waves are observed over a wide range of $C_0$ and $\kappa_1/\kappa_0$ (Fig. 2a). Figure 2b depicts snapshots of the pattern development for $C_0R = 4, \kappa_1/\kappa_0 = 4$, and $V^* = 0.8$. These snapshots demonstrate that propagating waves occur spontaneously from the two poles of a prolate-shaped vesicle, collide, and disappear at the center of the vesicle. In addition, propagating waves occur on the discocyte-shaped vesicles; for large values of $C_0$, they arise from the highly curved regions, whereas for $C_0 = 0$, the waves arise from the planar regions (Fig. 2c and d). The curvature effect of the term $K_{cv}$ increases in the curved regions for large values of $C_0$, whereas in the planar regions, it increases for small values of $C_0$. These nonuniform inputs, depending on $C_0$ and the vesicle shape, can induce propagating waves. Because of the axisymmetry of the vesicle shapes, the reaction-diffusion waves are also axisymmetric (Fig. 2b–d). However, for intermediate values of $C_0$, the curvature effect could not produce significant nonuniform inputs and lose axisymmetric waves (An example is shown for $C_0R = 1, \kappa_1/\kappa_0 = 2$, and $V^* = 0.65$, in Fig. S2a provided in the ESI). These results indicate that propagating waves spontaneously occur due to mechanochemical coupling, and their emergence is affected by the spontaneous curvature $C_0$ and the local membrane curvature.

In addition, unordered waves occur for larger $C_0$ and small $\kappa_1/\kappa_0$, exhibiting various shapes (green circles in Fig. 2a). For large values of $C_0$, the waves with long fronts are unstable and split (Fig. 2e); subsequently, the waves form spiral shapes and collide, leading to complicated spatiotemporal patterns. Furthermore, for large $D_u$, the unordered waves no longer propagate and change into nonpropagating patterns (Fig. 2f); spot or short stripe patterns are formed locally and stay at one position for a long time. For larger values of $D_u$, broader waves could be generated, allowing local membrane bending for large values of $C_0$ in the wave regions. Due to local membrane deformation, the spot and stripe patterns have large positive curvatures, whereas the surrounding regions have negative or small positive curvatures, resulting in negative feedback to the chemical reaction. Therefore, wave collision can be prevented by the surrounding regions, and spots and stripes appear. For $D_u = 4$, spot and stripe patterns are generated, but they are unstable (Fig. S2b provided in the ESI). A similar curvature effect of the reaction-diffusion patterns has been reported for a tissue deformation system[33]. Mechanochemical coupling induces the spontaneous occurrence of propagating waves, which affect pattern emergence and the development of reaction-diffusion waves depending on the spontaneous curvature, bending rigidity, and diffusion coefficients.

**Relationship between the wave speed and membrane curvature**
To understand the effect of membrane curvature on reaction-diffusion wave properties, we examined the wave speed on the deformable vesicles. We first calculated the wave speed $s_{wave}$ on the deformable discocyte (Fig. 2c), as shown in Fig. 3a (the algorithm for the calculation of $s_{wave}$ is presented in the ESI). For large values of $C_0$, $s_{wave}$ is correlated with the smoothed local membrane curvature $\widetilde{H}$. In addition, similar results are observed for the waves on the prolate vesicles (Fig. S4 provided in the ESI). For large values of $C_0$, the waves occur in the highly curved regions and propagate to the planar regions. To investigate whether membrane

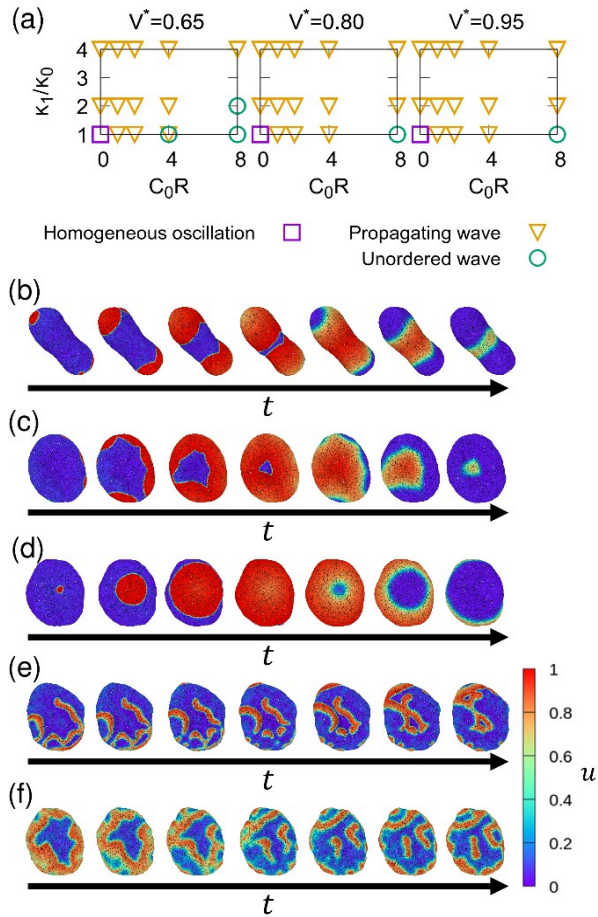

**Figure 2** (a) Pattern diagrams with deformable vesicles for $N = 15994$ and $V^* = 0.65, 0.8,$ and $0.95$. Prolate and discocyte-shaped vesicles are used as the initial states for (b) $V^* = 0.8$ and (c–f) $V^* = 0.65$, respectively. (b–f) Typical sequential snapshots for (b, c) $C_0R = 4$ and $\kappa_1/\kappa_0 = 4$, (d) $C_0R = 0$ and $\kappa_1/\kappa_0 = 4$, (e) $C_0R = 8$ and $\kappa_1/\kappa_0 = 2$ and (f) $C_0R = 8, \kappa_1/\kappa_0 = 1$, and $D_u = 16$. The color indicates the concentration of the curvature-inducing protein $u$. Each frame step is $1\tau$.

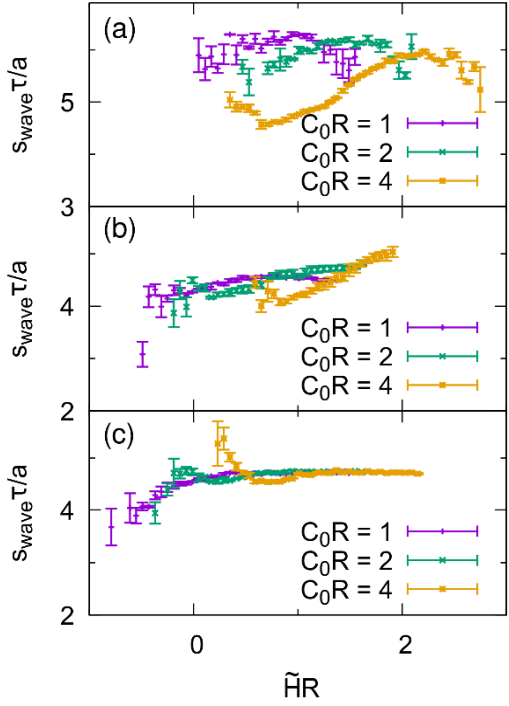

**Figure 3** Relationship between the membrane curvature and wave speed for discocyte-shaped vesicles ($V^* = 0.65$) at $N = 15994$ and $\kappa_1/\kappa_0 = 4$. (a) with mechanochemical feedback and without a stimulated vertex, (b) with mechanochemical feedback and a stimulated vertex and (c) without mechanochemical feedback and with a stimulated vertex. The corresponding snapshots are provided in the ESI (Fig. S5).

curvature affects the wave speed or the speed difference is due to the transient dynamics when the wave starts, we performed simulations with the stimulated vertex on the planar region of the oblate vesicle with and without mechanochemical feedback (Figs. 3b and c, and S5 in provided the ESI). On the stimulated vertex, an additional constant input of "1" is imposed on the right of eqn (4) for initiating the propagating waves. It is determined that the wave speed and membrane curvature do not correlate in the absence of mechanochemical feedback, whereas they correlate with mechanochemical feedback, even if the reaction-diffusion waves commence from the planar region (Fig. 3b and c). These results indicate that the wave speed is affected by the local membrane curvature through mechanochemical coupling.

For further investigation, we performed simulations with spherical vesicles of different sizes (i.e., $V^* = 1$) with the stimulated vertex. Without mechanochemical feedback, the wave speed correlates slightly with the membrane curvature because diffusion processes can be slightly affected by the curvature (Fig. S6 provided in the ESI). On the other hand, with mechanochemical feedback, it is found that the wave speed and membrane curvature are positively correlated for large values of $C_0$ and small values of $\kappa_1/\kappa_0$, whereas they are negatively correlated for small values of $C_0$ and large values of $\kappa_1/\kappa_0$ (Fig. 4a–c). This suggests that there is positive correlation when the binding of proteins to the membranes is energetically favorable, and negative correlation when it is energetically unfavorable.

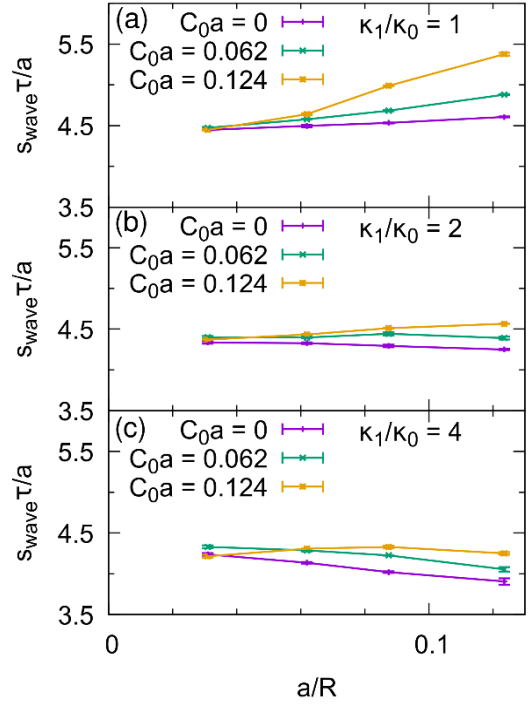

**Figure 4** Wave speed for spherical vesicles ($V^* = 1$) of four sizes ($N = 1006, 2004, 4000,$ and $15994$) with mechanochemical feedback and a stimulated vertex. The simulation conditions are at $C_0 a = 0, 0.062,$ and $0.124$, and (a) $\kappa_1/\kappa_0 = 1$, (b) $\kappa_1/\kappa_0 = 2$, and (c) $\kappa_1/\kappa_0 = 4$.

To confirm this concept, we analytically calculated the speed of a periodic traveling wave in one dimension, $r \in [0, L]$. According to a previous study[34, 35], a traveling wave with $\tilde{r} = r - ct$ and wavenumber $q = 2n\pi/L$ ($n \in \mathbb{N}$) is considered, where $c$ is the speed of the traveling wave, and the following equations are obtained:

$$\frac{\partial}{\partial t}\begin{pmatrix}\delta u \\ \delta v\end{pmatrix} = \mathbf{J}\begin{pmatrix}\delta u \\ \delta v\end{pmatrix},$$

$$\mathbf{J} = \begin{pmatrix} B - S_u - D_u q^2 & u_s^2 \\ -B & -u_s^2 - D_v q^2 \end{pmatrix},$$

$$S_u = \frac{u_s}{1 - u_s}\left(\frac{1}{1 - u_s} + 1\right)$$

For the above, $\text{tr}\mathbf{J} = 0$ yields the critical value of $B$ for instability. At $\text{tr}\mathbf{J} = 0$ and $\det\mathbf{J} > 0$, $\mathbf{J}$ has pure complex eigenvalues, and there is a traveling wave solution with a critical wave speed $c(q)$, which can be expressed as follows:

$$c(q)^2 = (D_u - D_v)u_s^2 + S_u\left(\frac{u_s}{q}\right)^2 - (D_v q)^2.$$

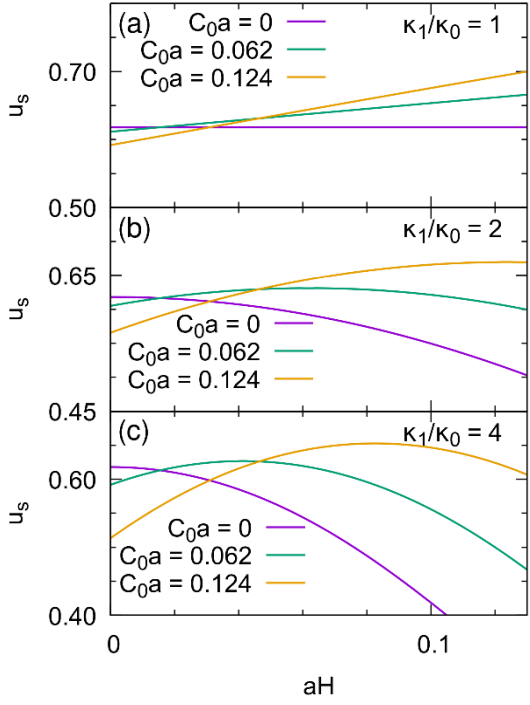

**Figure 5** Membrane curvature dependence of the fixed point of the curvature-inducing protein concentration $u_s$ for $C_0 a = 0, 0.062$ and $0.124$, and (a–c) $\kappa_1/\kappa_0 = 1, 2, 4$, respectively.

Further details on this analysis method can be found in ref. 34 and ref. 35. Because $c(q)$ monotonously increases with an increase in $u_s$ under the conditions adopted in this study ($D_u = D_v$), the membrane curvature dependency of the wave speed could be caused by that of $u_s$. Then, the membrane curvature modifies the speed through the fixed point of the curvature-inducing protein $u_s$ as follows:

$$\frac{\partial u_s}{\partial H} = [2(\kappa_0 - \kappa_1)H + \kappa_1 C_0]\frac{K_{cv}A}{k_B T}\left(\frac{1 + \frac{2}{K_{cv}A}}{\sqrt{1 + \frac{4}{K_{cv}A}}} - 1\right)$$

The sign of $\partial u_s/\partial H$ is determined by that of $[2(\kappa_0 - \kappa_1)H + \kappa_1 C_0]$. Therefore, for $\kappa_1/\kappa_0 = 1$, $u_s$ and $c(q)$ monotonously increase with an increase in $H$; in contrast, for $\kappa_1/\kappa_0 > 1$, $u_s$ and $c(q)$ monotonously increase for $H < H_{th} = \kappa_1 C_0/2(\kappa_1 - \kappa_0)$ and decrease for $H > H_{th}$. The dependence of $u_s$ on the membrane curvature is shown in Fig. 5. Although this analysis is for the critical wave speed of the periodic wave in one dimension, it captures the curvature dependence of the wave speed in the simulations with spherical vesicles (Figs. 4 and 5). This confirms the concept that when protein binding is energetically favorable (i.e., when $u_s$ becomes larger), the wave speed and membrane curvature are positively correlated. The wave speed and local membrane curvature correlate, and the relationship depends on the spontaneous curvature and bending rigidity.

**Self-oscillation of the vesicle shape**

Thus far, we have performed simulations under the conditions where the reaction-diffusion wave propagation was faster than membrane deformation. Further, we studied the interaction between membrane deformation and the reaction-diffusion waves with apparently comparable time scales, $\tau \gtrsim 100$ (see Fig. 6). Under these conditions, vesicle deformation is accompanied by reaction-diffusion waves. For $V^* = 0.65$, an oblate vesicle rapidly morphs into a prolate form, and a cyclic change in the vesicle shape occurs between the prolate form and two spheres connected by a small neck. For $V^* = 0.8$ and $0.95$, the vesicle shapes also oscillate; however, for $V^* = 0.95$ a closed neck is not formed (Fig. 6b and c). In addition, for four times smaller diffusion constant ($D_u = 1$) at $V^* = 0.65$ and $\eta = 1$, the waves do not cover the entire area of the small spheres (Fig. S7 provided in the ESI). As the membrane deformation propagates, the protein waves collide at the center and narrow the vesicle leading to the formation of a closed neck. Therefore, a broad wave to cover the whole area of budded spheres is not necessary to induce large shape oscillations.

Figure 7 shows the time development of the averaged concentration of $u$ over the vesicle and the integrated mean curvature $C_{IM} = (1/4\pi R)\int H dS$, which is the normalized area difference between the two monolayers of the bilayer membrane[32, 36, 37]. For a sphere, $C_{IM} = 1$. When the vesicle shape includes budded spheres connected by narrow necks, $C_{IM}$ distinctly increases (compare Figs. 6 and 7). Therefore, the vesicle shape oscillation, as shown in Fig. 6, is distinguished by calculating $C_{IM}$. The vesicles exhibit clearer oscillation at $V^* = 0.65$ compared to $V^* = 0.8$ and $0.95$. Starting from the prolate-shaped vesicles at $V^* = 0.65$, we performed simulations for various values of $\tau$ in order to analyze the relationship between the time scale and shape oscillation and determined that there was no apparent oscillation of $C_{IM}$ for a small $\tau$ whereas a large $\tau$ resulted in a large amplitude for $C_{IM}$ (Fig. 8). Thus, reaction-diffusion waves are also accompanied by large membrane deformation, depending on their time scale.

**Conclusions**

In this study, we have analyzed reaction-diffusion waves coupled with vesicle deformation. In the employed modified Brusselator model, spatially homogeneous oscillation in the protein concentration always occurs after relaxation at spontaneous curvature $C_0 = 0$ and bending rigidity ratio $\kappa_1/\kappa_0 = 1$ (i.e., there

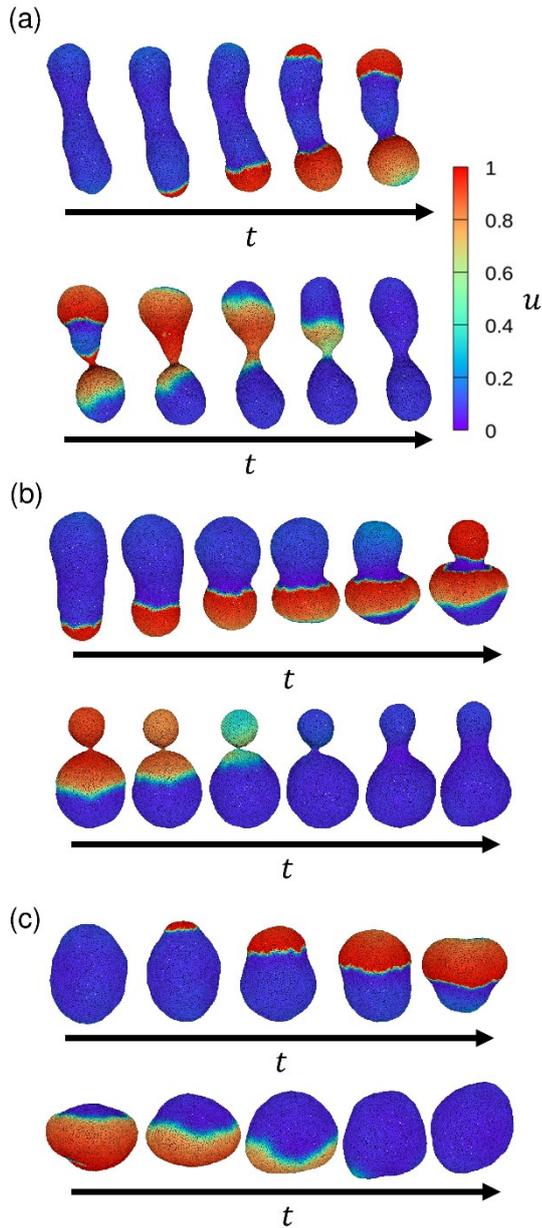

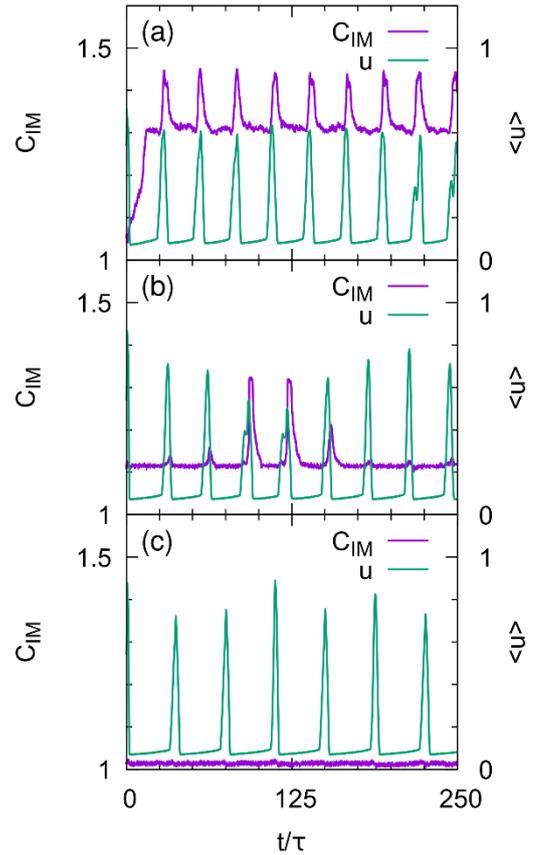

**Figure 6** Typical example of the self-oscillation of a vesicle shape due to coupling between the membrane curvature and reaction-diffusion system. Sequential snapshots of the shape oscillation for (a) $V^* = 0.65$, (b) $V^* = 0.8$, and (c) $V^* = 0.95$ at $N = 4000, \kappa_1/\kappa_0 = 4, C_0 R = 4, \tau = 400$. The color indicates the concentration of the curvature-inducing protein $u$. Each frame step is $1\tau$.

**Figure 7** Oscillations of the integrated mean curvature, $C_{IM}$ and the averaged value of the protein concentration, $\langle u \rangle$ for (a) $V^* = 0.65$, (b) $V^* = 0.8$, and (c) $V^* = 0.95$. The data are the same as in Fig. 6a–c, respectively

is no curvature effect on the chemical reactions). On the other hand, propagating waves spontaneously occur due to the curvature effect in the presence of mechanochemical coupling. The waves originate from a highly curved or flat region for a high or low $C_0$ value, respectively. Depending on $C_0$, $\kappa_1/\kappa_0$, and the vesicle shape, the propagating wave patterns change into various shapes, and even nonpropagating patterns appear. In addition, the membrane curvature affects the wave speed and exhibits positive or negative correlation depending on the magnitude of the spontaneous curvature and bending rigidity. This speed variation can be one of the sources for the formation of various spontaneous propagating-wave patterns on nonuniform surface shapes, such as the change to unordered waves on deformable vesicles. Modification of the reaction-diffusion patterns due to diffusion on a nonuniform curvature has been reported for fixed shapes[38, 39]. Our results demonstrated that the curvature effects are more pronounced through mechanochemical coupling and can induce shape deformation, in addition.

In a reconstituted Min system in GUVs, the vesicle size regulated the emergence of Min waves[40]. As Min proteins can induce membrane deformation[23, 24, 41], the relationship between the reaction-diffusion system and membrane curvature observed in this study may also be involved in the properties of such Min waves in GUVs. Moreover, for slow reaction-diffusion waves (large values of $\tau$), it is found that the vesicle shape is self-oscillating and associates with the reaction-diffusion waves of curvature-inducing proteins. This result agrees with the experimental observation in the Min system in GUVs[23-25]. In the theoretical analysis of ref. 25, the spontaneous curvature of the membrane was assumed to oscillate homogeneously on the entire surface. Our simulations revealed that change in the

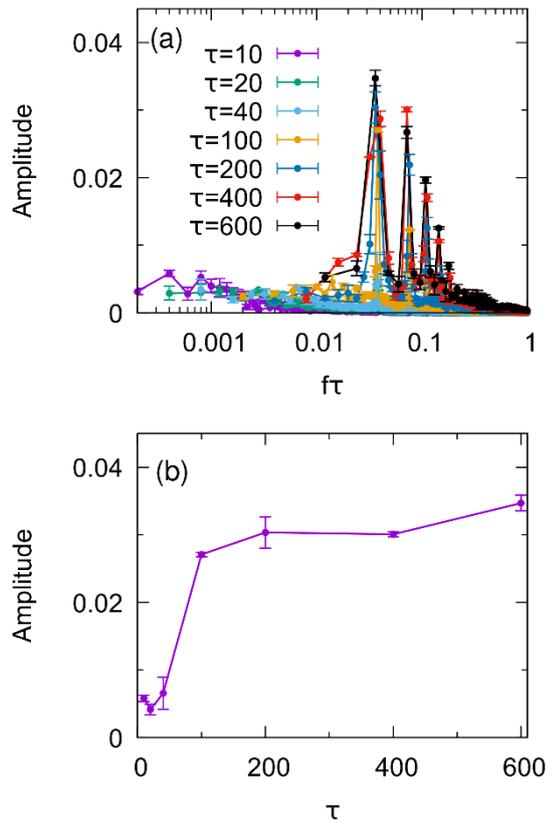

**Figure 8** (a) Fourier spectrum of $C_{IM}$ for $N = 4000, V^* = 0.65, \kappa_1/\kappa_0 = 4, C_0 R = 4$ and various values of $\tau$. (b) Maximum amplitude of the Fourier spectrum for various values of $\tau$.

spatial protein concentration accompanied by wave propagation could be directly reflected in the vesicle shape.

In addition to curvature-inducing proteins, actin filaments can change the membrane shape[13, 15, 21]. For example, in macropinocytosis, propagating waves of actin are observed, coupled with membrane deformation[14, 18]. Moreover, adherent cells, such as amoeboid and epithelial cells, exhibit reaction-diffusion waves on basal membranes, where the deformation is constrained by the substrates. In the experimentally observed waves in adherent cells, it has been reported that the waves mediated by membrane shape deformation had 10- to 100-fold faster speed compared to that of actin waves not associated with curvature-inducing proteins[13]. In our simulations, although the wave speed is affected by mechanochemical coupling, wave speeds with and without the curvature effect are of the same order of magnitude. Thus, the effects of actin filaments and substrate adhesion are important problems for further studies.

In living systems, the spatiotemporal oscillating events of curvature-inducing proteins on various shapes play important roles such as endocytosis[12], cell migration,[22] and cell division[8, 10, 24]. In future, we intend to explore the relationship between wave properties and biological functions.

## Conflicts of interest

There are no conflicts to declare.

## Acknowledgements

We would like to thank K. Fujiwara (Keio Univ.) and M. Yanagisawa, S. Ishihara, and R. Nishide (Univ. Tokyo) for stimulating discussions. This work was supported by JSPS KAKENHI Grant Number JP21K03481.

## Author Contributions

N.T. and H.N. designed the research. N.T. performed the computations and analyzed the data. N.T. and H.N. wrote the manuscript.

# Supplementary Materials: Reaction-Diffusion Waves Coupled with Membrane Curvature

*Naoki Tamemoto and Hiroshi Noguchi*[*]

Institute for Solid State Physics, University of Tokyo, Kashiwa, Chiba 277-8581, Japan

**SUPPLEMENTARY NOTES**

**Linear stability analysis**

The fixed point $(u_s, v_s)$ of reaction equations (2) and (3) is determined by solving $f(u,v) = 0$ and $g(u,v) = 0$:

$$u_s = \frac{2}{1 + \sqrt{1 + \frac{4}{A}}}, \quad v_s = \frac{B}{u_s}.$$

Thus, $u_s \to 0$ at $A \to 0$ and $u_s \to 1$ at $A \to \infty$. It is the crossing point of the two nullclines in Fig. S1a. For a small perturbation $(\delta u, \delta v)$ around the fixed point, we determine the equations of the first order in $\delta u$ and $\delta v$ as follow:

$$\frac{\partial}{\partial t}\begin{pmatrix}\delta u \\ \delta v\end{pmatrix} = \mathbf{J}\begin{pmatrix}\delta u \\ \delta v\end{pmatrix},$$

$$\mathbf{J} = \begin{pmatrix} B - S_u + D_u \nabla^2 & u_s^2 \\ -B & -u_s^2 + D_v \nabla^2 \end{pmatrix},$$

$$S_u = \frac{u_s}{1 - u_s}\left(\frac{1}{1 - u_s} + 1\right).$$

For a quasi-spherical vesicle, we obtain

$$\mathbf{J}_l = \begin{pmatrix} B - S_u - D_u \frac{l(l+1)}{r^2} & u_s^2 \\ -B & -u_s^2 - D_v \frac{l(l+1)}{r^2} \end{pmatrix},$$

based on the spherical harmonic expansion, where $r$ is the radius of the sphere and $-l(l+1)$ is an eigenvalue of the Laplace–Beltrami operator $\nabla^2$. In the present model, the conditions for the Hopf and Turing bifurcations are $B > S_u + u_s^2$ and $B >$

$\left(\sqrt{S_u} + u_s\eta\right)^2$, respectively, where $\eta = \sqrt{D_u/D_v}$. The phase diagram for $A = 1$ is presented in Fig. S1b.

**Potentials for molecular dynamics simulation**

For the molecular dynamics simulation in this study, we use the potential $U = U_S + U_V + U_b + U_r + U_{cv}$, where $U_S$ and $U_V$ are surface area and volume constraint potentials, $U_b$ and $U_r$ are bond and repulsive potentials, and $U_{cv}$ is a bending potential of membranes ($F_{cv}$ is discretized using dual lattices). The constraint potentials are written as below:

$$U_S = \frac{1}{2}k_S(S - S_0)^2,$$

$$U_V = \frac{1}{2}k_V(V - V_0)^2,$$

where $k_S$ and $k_V$ are constraint coefficients, $S_0$ and $V_0$ are references of surface area and volume, respectively. We use the following parameter values:

$$k_S = \frac{4k_B T}{a^2}, k_V = \frac{2k_B T}{a^3}, S_0 = 0.41(2N - 4), V_0 = \frac{4\pi R^3}{3}V^*.$$

For the bond and repulsive potentials, we employ a well-like potential, which has broad and flat bottom and exhibits a rapid increase to $\infty$ at $l_{c0} = 1.15a$ (bond) and at $l_{c1} = 0.85a$ (repulsion). More details of the potential and method are described in Refs. 28 and 29.

**Wave speed calculation method**

The wave speed is calculated based on the distance traveled by the wave front per $0.1\tau$. We determined that spatial waves exist when $u_{max} > 0.6$ and $u_{max} - u_{min} > 0.5$, where $u_{max}$ and $u_{min}$ are the maximum and minimum values of $u$ on the vesicle, respectively. When waves exist, we identify the vertices for $u > 0.6$ as belonging to the wave region and the vertices adjacent to the nonwave regions as the wave edges (i.e., vertices on the inner boundaries between the wave regions and nonwave regions). The front edge is determined by the condition of $v$ at the vertices, $v \geq v_{min} + 0.4(v_{max} - v_{min})$, where $v_{max}$ and $v_{min}$ are the maximum and minimum values of $v$ on the vesicle, respectively, because when the reaction-diffusion waves propagate, the oscillation phase of $v$ should shift from that of $u$ toward the propagation direction (Fig. S2). If the wave topology does not change (i.e., when the waves do not merge or split) and the wave front has more than 10 vertices, we calculate the minimum displacement from each vertex to the vertices at the edge of the wave at the previous or next coordinates (vertex coordinates at time $\pm 0.1\tau$). If the standard deviation of the minimum displacement for each vertex is greater than $0.5a$, the calculation is rejected, if not, we calculate the average displacement as the distance traveled by the wave front.

# SUPPLEMENTARY FIGURES

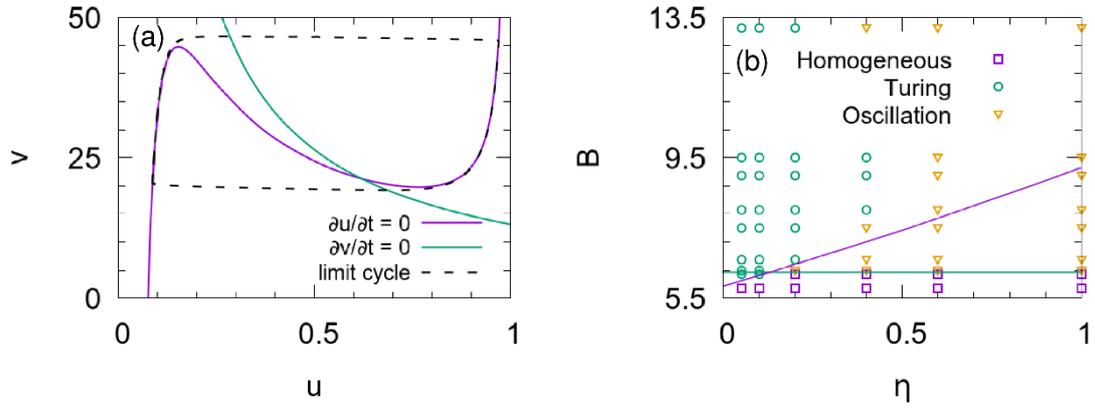

Fig. S1 (a) Phase plane of the Brusselator model with the modified protein unbinding process for $A = 1$ and $B = 13.2$. (b) Phase diagram for a spherical vesicle ($V^* = 1$) at $A = 1$ and $D_u = 4$ in the absence of membrane curvature feedback to the Brusselator. The purple and green lines on the phase diagram represent the Turing and Hopf bifurcation curves, respectively. The symbols represent the simulation results.

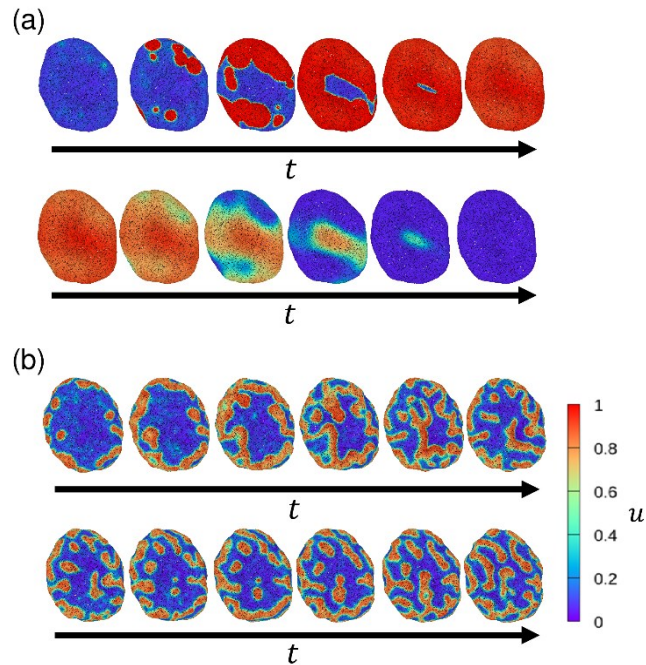

Fig. S2 Example snapshots of the pattern formation on discocyte-shaped vesicles ($V^* = 0.65$) for (a) $C_0 R = 1, \kappa_1/\kappa_0 = 2$, and $D_u = 4$, and (b) $C_0 R = 8, \kappa_1/\kappa_0 = 1$, and $D_u = 4$. The color indicates the concentration of the curvature-inducing protein $u$. Each frame step is $0.5\tau$ for (a) and $2\tau$ for (b).

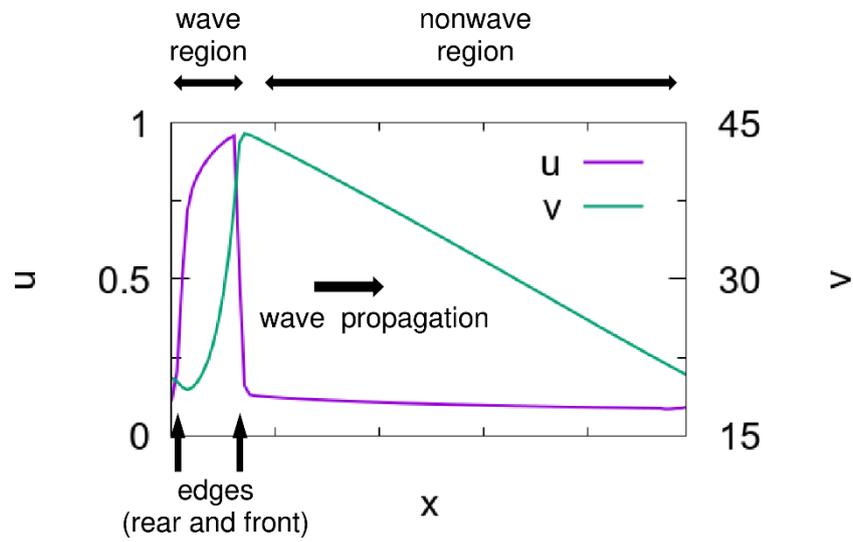

Fig. S3 Protein concentration profiles in a propagating wave in one-dimensional space. The front and rear wave edges have different concentrations of $v$ because of the phase shift between $u$ and $v$.

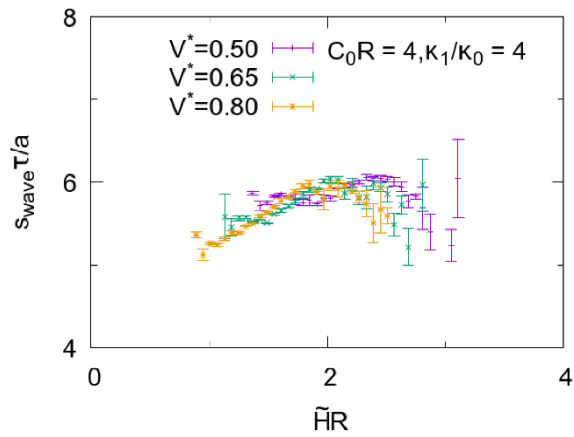

Fig. S4 Relationship between the membrane curvature and wave speed at $\kappa_1/\kappa_0 = 4, C_0 R = 4, D_u = 1,$ and $N = 15994$ for prolate shapes with $V^* = 0.5, 0.65,$ and $0.8$.

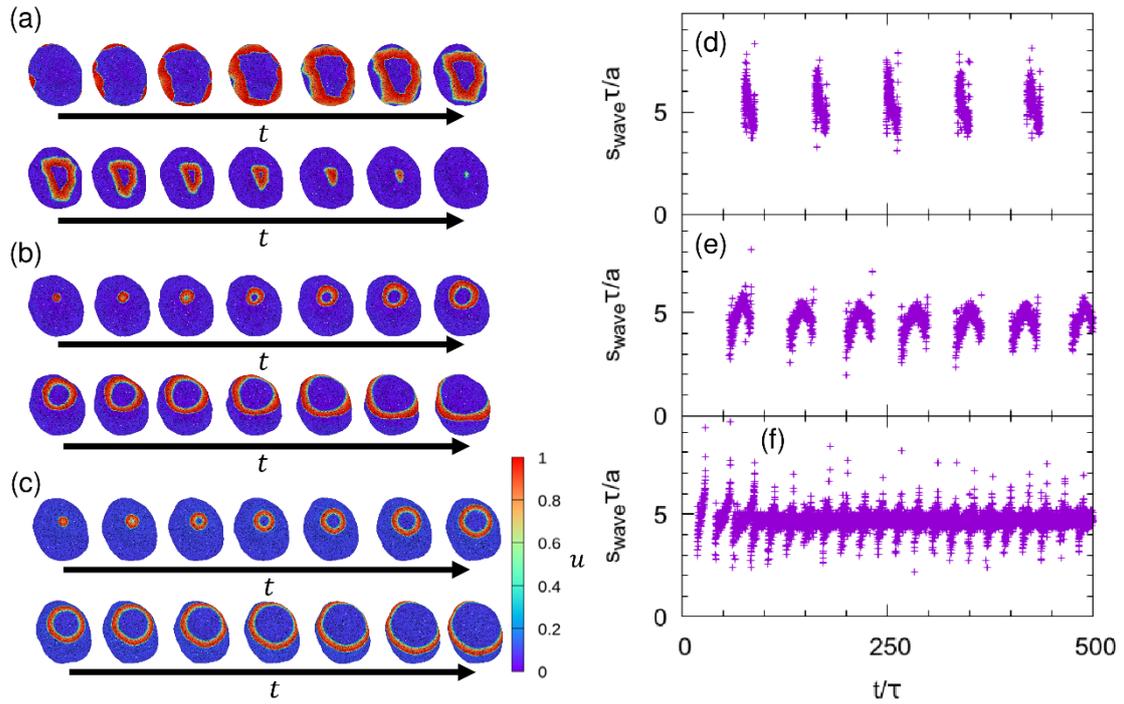

Fig. S5 (a–c) Typical sequential snapshots of the propagating reaction-diffusion waves for $N = 15994, V^* = 0.65, \kappa_1/\kappa_0 = 4, C_0 R = 4,$ and $D_u = 1$ (a) with mechanochemical feedback and without a stimulated vertex, (b) with mechanochemical feedback and a stimulated vertex, and (c) without mechanochemical feedback and with a stimulated vertex. The color indicates the concentration of the curvature-inducing protein, $u$. Each frame step is $1\tau$. (d–f) Time development of the wave speed under conditions corresponding to (a–c), respectively.

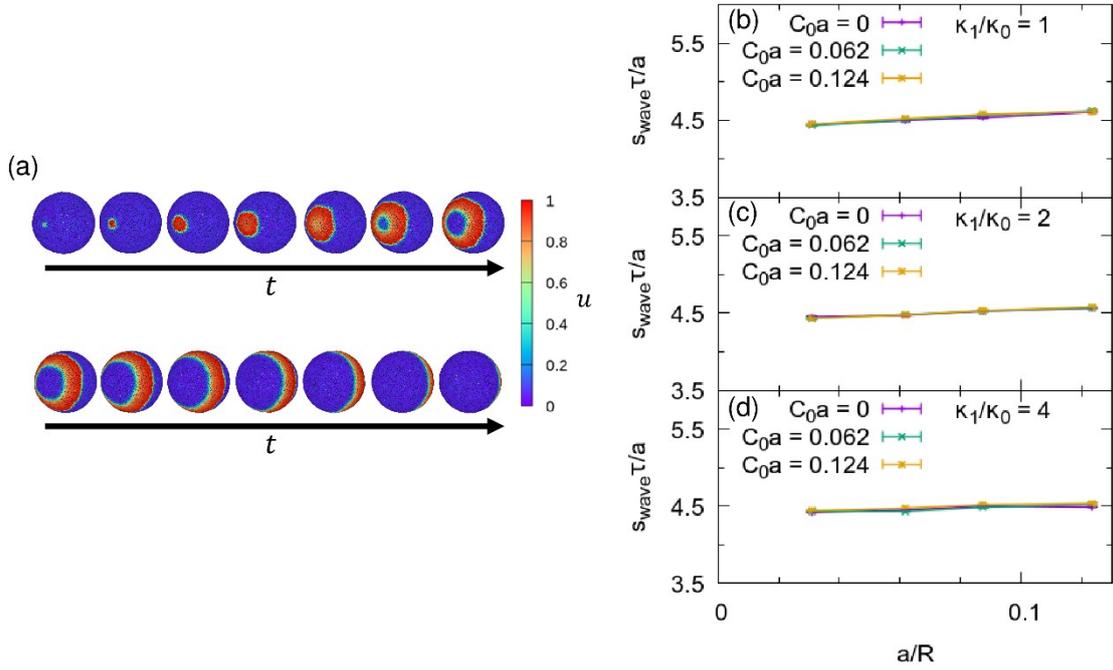

Fig. S6 (a) Example snapshots of the propagating waves on a spherical vesicle for $N = 4000, C_0 a = 0.124 \ (C_0 R = 2)$, and $\kappa_1/\kappa_0 = 4$. Each frame step is $1\tau$. (b–d) Wave speed for $N = 1006, 2004, 4000, 15994$ and $V^* = 1$ with a stimulated vertex and without mechanochemical feedback for $\kappa_1/\kappa_0 = 1, 2, 4$, respectively.

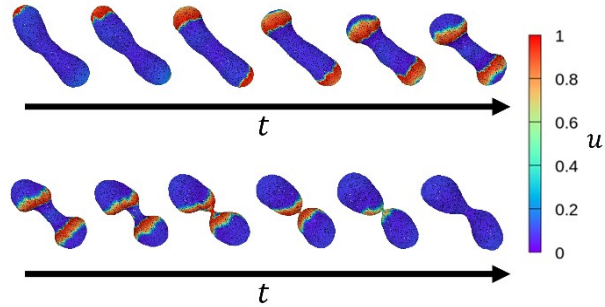

Fig. S7 Self-oscillation of a vesicle shape due to coupling between the membrane curvature and reaction-diffusion system for $N = 4000, \kappa_1/\kappa_0 = 4, C_0 R = 4, D_u = 1, \tau = 400$, and $V^* = 0.65$. Sequential snapshots of the shape oscillation. The color indicates the concentration of the curvature-inducing protein $u$. Each frame step is $1.25\tau$.